\documentclass[aps,pre,twocolumn,groupedaddress,showpacs]{revtex4-1}

\usepackage{natbib}
\usepackage{graphics}
\usepackage{amsfonts}
\usepackage{amssymb}
\usepackage{amsmath}

\begin{document}

\title{Hard ellipsoids: analytically approaching the exact overlap distance}

\author{F. de J. Guevara-Rodr\'{\i}guez}
\author{G. Odriozola}
\email{Corresponding author email: godriozo@imp.mx} 
\affiliation{Programa de Ingenier\'{\i}a
Molecular, Instituto Mexicano del Petr\'{o}leo, Eje Central
L\'{a}zaro C\'ardenas 152, 07730, M\'{e}xico, Distrito Federal,
M\'{e}xico.}

\date{\today}

\begin{abstract}
Following previous work (JCP 134, 201103 (2011)), the replica exchange Monte Carlo technique is used to produce the equation of state of hard 1:5 aspect-ratio oblate ellipsoids for a wide density range. Here, in addition to the analytical approximation of the overlap distance given by Berne and Pechukas (BP) and the exact numerical solution of Perram and Wertheim, we tested a simple modification of the original BP approximation (MBP) which corrects the known T-shape mismatch of BP for all aspect-ratios. We found that the MBP equation of state shows a very good quantitative agreement with the exact solution. The MBP analytical expression allowed us to study size effects on the previously reported results. For the thermodynamic limit, we estimated the exact 1:5 hard ellipsoid isotropic-nematic transition at the volume fraction $0.343 \pm 0.003$, and the nematic-solid transition in the volume fraction interval $(0.592 \pm 0.006)-(0.634 \pm 0.008)$. 
\end{abstract}


\maketitle

\section{Introduction}

In 1972 Berne and Pechukas~\cite{Berne72} (BP) published a pioneer work where molecules were represented by an uniaxial ellipsoidal Gaussian and their pair interaction was then related to their overlap integral. This was probably one of the first attempts to produce a coarse-grained model and it certainly was a keystone for the development of the widely used anisotropic Gay-Berne~\cite{Gay81} interaction. This last pair-potential is very popular since it allows capturing the orientational ordering of the entities in nematic and/or smectic phases while being a simple analytical expression. Hence, it has become one of the first choices  to model liquid-crystals~\cite{McGroder98,Brown98,deMiguel02,deMiguel04,Care05,Wilson05,Wittkowski10,Jimenez-Serratos11}.

In the introduction of the BP work it is said: \emph{The potential must have two characteristics: It must be mathematically simple, involving only functions which are easy to calculate; and it must not violate too strongly our sense of what is physically correct.} Almost forty years later, with the huge advance of computing power, the development of very efficient and easy to use software based on the so called force-fields, and the possibility of simulating a complete and solvated protein by handling several thousands of atoms, one is tempted to think that these lines are obsolete. Notwithstanding, there is nowadays a great effort on developing and studying pair potentials which goes in the coarse-grained direction~\cite{Dijkstra97,Odriozola04,Detcheverry08,Patti11,Paramonov05}. In fact, ellipsoids are frequently used as models for highly anisotropic and relatively rigid supramolecular structures, such as biological membranes, DNA-phospholipidic complexes, and other anisotropic entities like nanotubes, inorganic nanorods, clay crystals, among a wide variety of colloidal particles. Additionally, and hardly contrasting with the huge advance of computing power, even for the most simple of the pair potentials, i.~e. for hard spheres, there are questions that still cannot be answered~\cite{Parisi10,Odriozola11}. Thus, it seems that these BP lines are not only current these days, but also they will remain valid for many more decades.

An ellipsoidal shape can be handled to match the shape of many molecules and colloids. For this purpose, however, it is desirable to know the exact shape of the model hard core, as well as its volume and surface. Unfortunately, the BP hard potential does not provide a defined shape and volume. On the other hand, the exact ellipsoidal hard core interaction is only numerically solvable~\cite{Perram84,Perram85,Paramonov05}. Thus, the development of better analytical approaches to the exact solution is convenient. In this regard, Rickayzen~\cite{Rickayzen98} proposed a modification of the original BP analytical expression (MBP) aiming to correct the well-known T-shape mismatch of BP. In this work, we slightly modify the Rickayzen expression (adding a fitting parameter) to approach even further the exact solution. The resulting expression presents a good balance between precision and complexity, and so, it reasonably fulfills the lines written by BP~\cite{Berne72}. Then, we tested the analytical expression against the exact solution by comparing the corresponding equations of states for particles with 1:5 aspect-ratio. For this purpose, we implemented the replica exchange Monte Carlo simulation method. We observed a very good agreement between the exact and the MBP analytical equations of state. The MBP approach allowed us to study size effects on the results reported in ref.~\cite{OdriozolaHE} Thus, for the thermodynamic limit, we estimate the occurrence of the exact 1:5 hard ellipsoid isotropic-nematic transition at the volume fraction $0.343 \pm 0.003$, and the fluid-solid transition in the volume fraction interval $(0.592 \pm 0.006)-(0.634 \pm 0.008)$. Finally, another crystal structure was captured in coexistence with that found in previous work~\cite{OdriozolaHE}.

The paper is organized as follows. The hard ellipsoidal pair potential models are given in the next section. There we compare the BP and the modified BP predictions with those provided by the exact numerical solution. Section III describes the replica exchange method for hard bodies. The equation of state and the structure of the studied systems are given in Section IV. Finally, we tackle the conclusions in Section V.

\section{Hard ellipsoidal models}

In this section we summarizes two useful tools to deal with hard ellipsoidal interactions. These are the Berne and Pechukas analytical approximation and an algorithm to produce the exact solution for a given precision. The following two subsections present these methods. A third subsection is devoted to introduce a modification of the Berne and Pechukas analytical expression which improves its overall performance. The improvement is shown in a final subsection.  

\subsection{Berne and Pechukas}

The Berne and Pechukas potential (BP)~\cite{Berne72} is a reasonable approximation for the exact interaction of two equal hard ellipsoidal particles. It is analytical, mathematically simple, easy to implement, fast to compute, and it can be used to study the condensed phase of a collection of prolate or oblate particles via numerical experiments~\cite{Ghoufi11,Luckhurst10,Berardi09,Luckhurst93}. In this approach, molecules are represented with an uniaxial ellipsoidal Gaussian and their interaction is then related to their overlap integral. In this way, a coarse-grained potential is built which successfully captures the anisotropic nature of the entities. The expression for the distance between the geometric centers of the ellipsoids when the particles are at contact, $\sigma_{BP}( \hat{\mathbf{u}}_{i}, \hat{\mathbf{u}}_{j}, \hat{\mathbf{r}} )$, is given by
\begin{widetext}
\begin{equation} \label{BP1}
\sigma_{BP}( \hat{\mathbf{u}}_{i}, \hat{\mathbf{u}}_{j}, \hat{\mathbf{r}} ) = \sigma_{\perp} \bigg (\ 1 - \frac{1}{2} \chi \bigg [\ \frac{ ( \hat{\mathbf{r}} \cdot \hat{\mathbf{u}}_{i} + \hat{\mathbf{r}} \cdot \hat{\mathbf{u}}_{j} )^{2} }{ 1 + \chi \hat{\mathbf{u}}_{i} \cdot \hat{\mathbf{u}}_{j} } + \frac{ ( \hat{\mathbf{r}} \cdot \hat{\mathbf{u}}_{i} - \hat{\mathbf{r}} \cdot \hat{\mathbf{u}}_{j} )^{2} }{ 1 - \chi \hat{\mathbf{u}}_{i} \cdot \hat{\mathbf{u}}_{j} } \ \bigg ] \ \bigg )^{-1/2}
\end{equation}
\end{widetext}
where $\hat{\mathbf{u}}_{i}$ and $\hat{\mathbf{u}}_{j}$ are the versors (unit vectors) along the main axis of each particle, and $\hat{\mathbf{r}}$ is the versor along the line joining the geometric centers. Here, the anisotropy parameter $\chi$ is 
\begin{equation}\label{BP2}
\chi = \frac{ \sigma_{\|}^{2} - \sigma_{\bot}^{2} }{ \sigma_{\|}^{2} + \sigma_{\bot}^{2} }
\end{equation}
where $\sigma_{\|}$ and $\sigma_{\bot}$ are the parallel and perpendicular diameters with respect to the main axis, respectively.

\subsection{Exact numerical solution}

An ellipsoidal surface centered at $\mathbf{r}_{i}$ and oriented according to $\hat{\mathbf{u}}_{i}$ (parallel to its main axis) is given by the following quadratic form
\begin{equation}\label{Ex1}
\mathcal{A}_{i}( \mathbf{r}_e ) = (\ \mathbf{r}_e - \mathbf{r}_{i}\ )^{t} \cdot \mathbb{A}_{i} \cdot (\ \mathbf{r}_e - \mathbf{r}_{i}\ ),
\end{equation}
where the scalar $\mathcal{A}_{i}( \mathbf{r} ) = const.$ and $\mathbf{r}_e$ is a point at the surface. In particular, the ellipsoid surface is represented by $\mathcal{A}_{i}( \mathbf{r}_e ) = 1$, having  
\begin{equation}\label{Ex2}
\mathbb{A}_{i} = \mathbb{U}^{t}( \hat{\mathbf{u}}_{i} ) \cdot \mathbb{D}^{-2} \cdot \mathbb{U}( \hat{\mathbf{u}}_{i} ),
\end{equation}
the rotation matrix $\mathbb{U}( \hat{\mathbf{u}}_{i} )$ (the matrix which converts a vector from the space-fixed to the body-fixed coordinate system~\cite{Goldstein}), its transpose $\mathbb{U}^{t}$, and the diagonal matrix $\mathbb{D}$,
\begin{equation}\label{Ex3}
\mathbb{D} = \frac{1}{2}\left(
\begin{array}{ccc}
\sigma_{\bot} & 0 & 0 \\
0 & \sigma_{\bot} & 0 \\
0 & 0 & \sigma_{\|} \\
\end{array}
\right).
\end{equation}
Thus, the geometry of the particle is given by $\mathbb{D}$ and its orientation is given by $\mathbb{U}( \hat{\mathbf{u}}_{i} )$ (or simply by $\hat{\mathbf{u}}_{i}$).

Let's consider two equal, arbitrarily oriented, and non-overlapping ellipsoids $i$ and $j$, at contact at point $\mathbf{r}_{c}$. The vector normal to the $i$ surface at the contact point is
\begin{equation}\label{Ex4}
\mathbf{n}_{i}( \mathbf{r}_{c} ) = \nabla \mathcal{A}_{i}( \mathbf{r}_{c} ) = \mathbb{A}_{i} \cdot (\ \mathbf{r}_{c} - \mathbf{r}_{i}\ ).
\end{equation}
A similar equation can be written for the vector normal to the $j$ surface $\mathbf{n}_{j}$. Since the tangent plane is common for both ellipsoids, the normal versors $\hat{\mathbf{n}}_{i}=\mathbf{n}_{i}/|\mathbf{n}_{i}|$ and $\hat{\mathbf{n}}_{j}=\mathbf{n}_{j}/|\mathbf{n}_{j}|$ fulfill
\begin{equation}\label{Ex5}
\hat{\mathbf{n}}_{i}( \mathbf{r}_{c} ) + \hat{\mathbf{n}}_{j}( \mathbf{r}_{c} ) = \mathbf{0}.
\end{equation}
This condition was originally employed by Perram and Wertheim for developing an algorithm for numerically determining the point $\mathbf{r}_c$~\cite{Perram84,Perram85}. In their work the Elliptic Contact Function (ECF) is introduced, which contains the information given by equation (\ref{Ex5}) and allows determining the distance of closest approach. Later, the ECF procedure is reviewed by Paramonov and Yaliraki, who contributed with a clear geometric interpretation of the Perram and Wertheim approach~\cite{Paramonov05}. The expression for the function that connects the particles centers through the geometric place where the vectors $\nabla \mathcal{A}_{i}( \mathbf{x}_{c} )$ and $\nabla \mathcal{A}_{j}( \mathbf{x}_{c} )$ are antiparallel is given by~\cite{Paramonov05}
\begin{equation}\label{Ex6}
\mathbf{x}_{c}( \lambda ) = \big ( \lambda \mathbb{A}_{i} + ( 1 - \lambda ) \mathbb{A}_{j} \big )^{-1} \cdot \big ( \lambda \mathbb{A}_{i} \cdot \mathbf{r}_{i} + ( 1 - \lambda ) \mathbb{A}_{j} \cdot \mathbf{r}_{j} \big )
\end{equation}
where $\lambda \in [0,1]$ is a scalar parameter. Note that for $\lambda=1$ and 0 the geometric centers of the ellipsoids $i$ and $j$ are obtained, respectively. The contact point $\mathbf{r}_c$ lies on this trajectory and corresponds to a unique value of $\lambda$, $\lambda_{c}$, which fulfills $\lambda_{c} \in (0,1)$. Furthermore, $\mathcal{A}_{i}\big (\mathbf{r}_c \big ) = \mathcal{A}_{j}\big ( \mathbf{r}_c \big )$, with $\mathbf{r}_c= \mathbf{x}_{c}( \lambda_{c})$.

With the above expressions it is easy to implement an iterative procedure to yield $\mathbf{r}_c$ with the desired precision. In particular, we implemented the bisection algorithm. It starts by evaluating $\Delta(\lambda)=\mathcal{A}_{i} (\mathbf{x}_c(\lambda)) - \mathcal{A}_{j} (\mathbf{x}_c(\lambda))$ for $\lambda=0.5$ ($\Delta(\lambda)$ is a monotonously decreasing function and is zero solely at $\lambda_{c}$). A positive $\Delta(\lambda)$ means $1>\lambda_c>\lambda$ and so, $\lambda$ is increased in such a way to reduce in half the interval. Conversely, a negative $\Delta(\lambda)$ means $0<\lambda_c<\lambda$ and $\lambda$ is decreased accordingly. In this way the interval is reduced as $1/2^n$, being $n$ the number of iterations. This method is simple and safe but not the faster. Approximately 20 iterations yields an error of $\Delta$ smaller that $1 \times 10^{-6}$. Note that the involved operations are only products and summations which translate into a fast computation. 

Additionally, the contact parameter $\lambda_{c}$ defines the extreme value of the linear combination of the quadratic forms $\mathcal{A}_{i}( \mathbf{x}_c )$ and $\mathcal{A}_{j}( \mathbf{x}_c )$, namely
\begin{equation}\label{Ex7}
\mathcal{S}( \lambda ) = \lambda\ \mathcal{A}_{i}\big ( \mathbf{x}_{c}( \lambda ) \big ) + ( 1 - \lambda ) \mathcal{A}_{j}\big ( \mathbf{x}_{c}( \lambda ) \big ).
\end{equation}
In other words, $0 \leq \mathcal{S}( \lambda ) \leq \mathcal{S}( \lambda_{c} )$ ($\mathcal{S}( \lambda )$ is an strictly concave function which ensures it has a unique maximum at $\lambda_{c}$). Note that from equations (\ref{Ex1}), (\ref{Ex2}), and (\ref{Ex7}), $\mathcal{S}( 0 ) = \mathcal{S}( 1 ) = 0$. The contact parameter also defines the Perram-Wertheim (PR) contact distance $\sigma_{PW}$,
\begin{equation}\label{Ex8}
\sigma_{PW} = \frac{r}{ \sqrt{ \mathcal{S}( \lambda_{c} ) } }.
\end{equation}
Hence, ellipsoids having their geometric centers separated a distance $r$ smaller than $\sigma_{PW}$ overlap whereas they do not for $r>\sigma_{PW}$. In other words, $(S(\lambda_c))^{-1/2}$ can be interpreted as a scaling factor needed to bring the particles into contact. In particular, the BP analytical expression for the contact distance corresponds to~\cite{Paramonov05}
\begin{equation}\label{Ex9}
\sigma_{BP} = \frac{r}{\sqrt{\mathcal{S}(1/2)}}
\end{equation}
which, in general, overestimates the exact result, $\sigma_{BP} \geq \sigma_{PW}$, since $\mathcal{S}(\lambda_c) \geq \mathcal{S}(1/2)$. 

\subsection{Modified Berne and Pechukas\label{subsecMBP}} 

\begin{figure*}
\resizebox{0.98\textwidth}{!}{\includegraphics{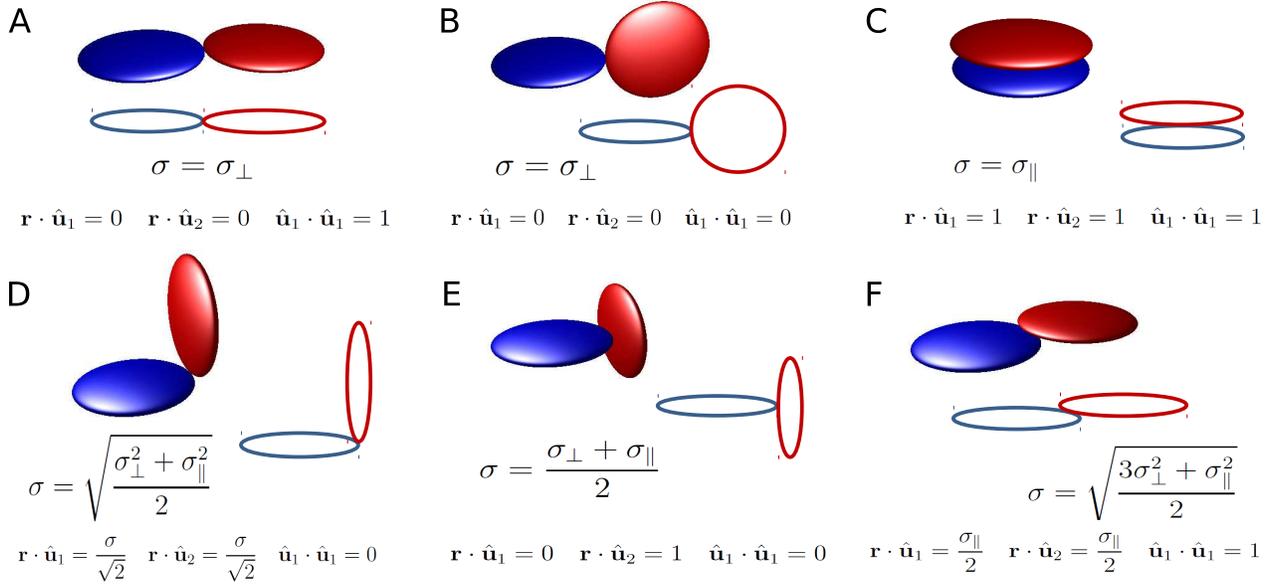}} \caption{\label{Fig1} Two ellipsoids at different contact configurations. The exact center-center distance at contact, $\sigma$, is given as a function of $\sigma_{\|}$ and $\sigma_{\bot}$. The relations among the normal versors $\hat{\mathbf{u}}_{i}$, $\hat{\mathbf{u}}_{j}$ and the center-center vector $\mathbf{r}$ are also shown. The BP approximation produces the exact solution for all cases but E. MBP yields the exact solution for all shown cases. } 
\end{figure*}

The BP approximation to the contact distance of two equal ellipsoids becomes poor for $\hat{\mathbf{u}}_{i}\hat{\mathbf{u}}_{j}\rightarrow0$. In particular, for case E of Fig.~\ref{Fig1} the BP result is $((\sigma_{\bot}^2+\sigma_{\|}^2)/2)^{1/2}$ instead of $(\sigma_{\bot}+\sigma_{\|})/2$, which is the exact solution. To improve the overall behavior of the BP approximation Rickayzen proposed~\cite{Rickayzen98}
\begin{equation}\label{MBP1}
\sigma_{MBP} = \sigma_{\bot} \bigg ( 1 - \frac{1}{2} \chi \big [ A^{+} + A^{-} \big ] + \big ( 1 - \chi ) \chi' \big [ A^{+} A^{-} \big ]^{\gamma} \bigg )^{-1/2}
\end{equation}
being
\begin{equation}\label{MBP2}
A^{\pm} = \frac{ ( \hat{\mathbf{r}} \cdot \hat{\mathbf{u}}_{i} \pm \hat{\mathbf{r}} \cdot \hat{\mathbf{u}}_{j} )^{2} }{ 1 \pm \chi \hat{\mathbf{u}}_{i} \cdot \hat{\mathbf{u}}_{j} }
\end{equation}
and
\begin{equation}\label{MBP3}
\chi' = \bigg ( \frac{ \sigma_{\|} - \sigma_{\bot} }{ \sigma_{\|} + \sigma_{\bot} } \bigg )^{2}.
\end{equation}
In his work $\gamma$ is unity. These expressions lead to very good results, since the introduced term does not alter the correct answers given by BP for cases A-D and F of Fig.~\ref{Fig1} (note that in these cases $A^{+} A^{-}=0$). Additionally, $\chi'$ is given such that case E is also satisfied. Thus, the above expressions correctly describe all cases shown in Fig.~\ref{Fig1}. Also note that $\sigma_{MBP} \leq \sigma_{BP}$ for all aspect ratios and $\gamma$.   

There are, however, other expressions which produce the correct result of cases A-F of Fig.~\ref{Fig1}. After trying some of them, we concluded that expressions (\ref{MBP1})-(\ref{MBP3}) (MBP) show a good compromise between accuracy and complexity. We solely added $\gamma$ as a free parameter to be adjusted as a function of the aspect ratio.   

\subsection{Comparing BP and MBP to the exact numerical solution}

\begin{figure}
\resizebox{0.45\textwidth}{!}{\includegraphics{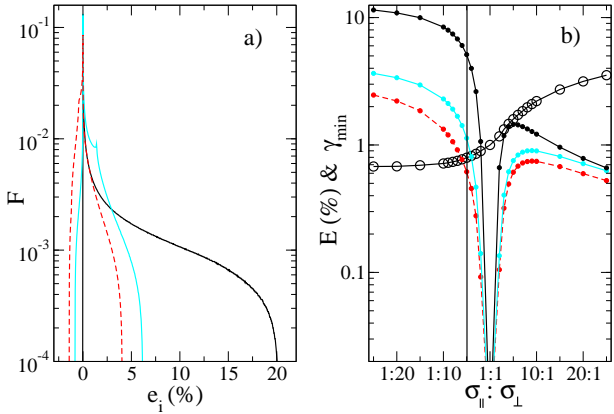}}
\caption{\label{Fig2} a) Appearance frequency, F, against the percentage of deviation of the analytical overlap distance with the corresponding exact distance, $e_i=100(\sigma_{an_i}-\sigma_{PW_i})/\sigma_{PW_i}$, for the 1:5 oblate case. The histogram is built by considering $N_c=1\times10^8$ random positions and orientations. b) Average deviation, $E=1/N_c\sum_i^{N_c}|e_i|$ (bullets), and the optimized $\gamma$ parameter, $\gamma_{min}$ (open symbols), for several aspect ratios. The vertical dark line at 1:5 points out the case shown in panel a). Black solid lines, cyan solid lines, and red dashed lines correspond to the BP, MBP with $\gamma=1$, and MBP with $\gamma_{min}$ analytical expressions, respectively. }
\end{figure}

In the previous section a modification of the BP (MBP) is presented, which fixes the T-shape mismatch while preserving the goodness of the BP analytical solution for the rest of the cases. We now compare the BP and the MBP predictions with the corresponding exact numerical solution (the error of the numerical solution is always below $10^{-8}$). This is shown in detail for the 1:5 oblate case in Fig.~\ref{Fig2} a). This plot shows the appearance frequency of a given deviation percentage as a function of the deviation percentage, $e_i=100(\sigma_{an_i}-\sigma_{PW_i})/\sigma_{PW_i}$. Here, $\sigma_{an_i}$ is the analytical prediction of the overlap distance for the random position and orientation $i$, and $\sigma_{PW_i}$ is the corresponding exact solution. The plot is then built by considering $N_c=1\times10^8$ different positions and orientations (note that for a spatially fixed ellipsoid we must generate a random position, keeping $r$ fixed, and a random orientation of the second ellipsoid to sample from the space of all possible configurations). The black, cyan, and dashed red lines represented in the plot correspond to the BP and the MBP results with $\gamma=1$ and $\gamma_{min}$, respectively. Here, $\gamma_{min}$ is the value of $\gamma$ which minimizes $E=1/N_c\sum_i^{N_c}|e_i|$. As can be seen, the BP deviations are greater or equal than zero, meaning that BP overestimates the volume of the corresponding oblate ellipsoid. Additionally, a large peak at zero is seen, which indicates that many configurations are correctly described. In the same direction, we should mention that 50$\%$ of the randomly generated configurations have a deviation smaller than 3.71$\%$. Nonetheless, the distribution also shows a long tail reaching deviations as large as 20$\%$. A look to the corresponding snapshots for large deviations shows that the ellipsoids are close to the T-shape configuration (not shown).  

Fig.~\ref{Fig2} a) shows the main goal of the MBP, i.~e., the long tail towards positive deviations of the BP expression is strongly suppressed. This occurs for both, $\gamma=1$ and $\gamma=\gamma_{min}$. That is, for all considered cases, we did not detect deviations larger than 6.04$\%$ and 4.04$\%$, respectively. From the simulation point of view this is a nice result since it guaranties configurations with $r<0.959\sigma_{MBP}$ to be overlapped. In the same line of thinking, configurations having $r>\sigma_{BP}$ cannot be overlapped. These results aid reducing the number of configurations to numerically solve when the purpose of the simulation is to work with the exact hard core model of the ellipsoids. 

Fig.~\ref{Fig2} a) also shows that a price must be paid when correcting the BP T-shape mismatch. That is, MBP introduces negative deviations. Consequently, there are overlapped configurations that MBP will consider as free of overlaps. However, we did not detect these deviations to exceed $0.84\%$ and $1.39\%$ for $\gamma=1$ and $\gamma=\gamma_{min}$, respectively. Thus, ellipsoids are definitely not overlapped for configurations having $r>1.020\sigma_{MBP}$. On the other hand, the average deviation, $E=1/N_c\sum_i^{N_c}|e_i|$, is reduced from 5.41 to 1.13 ($\gamma=1$) and to 0.62 ($\gamma=\gamma_{min}$) when implementing MBP instead of BP, which points out that the true ellipsoidal shape is much better approached. 

In brief, Fig.~\ref{Fig2} a) shows an improvement of MBP when compared to BP for the 1:5 aspect-ratio oblate case. Now, what happens when considering other aspect ratios or prolate ellipsoids? The answer to this question is given in Fig.~\ref{Fig2} b). There, the average deviation, $E=1/N_c\sum_i^{N_c}|e_i|$, as a function of the aspect ratio is shown for both, oblate and prolate ellipsoids. Again, the black line is used to represent the BP results whereas the cyan line and the red dashed line correspond to the MBP expression with $\gamma=1$ and $\gamma=\gamma_{min}$, respectively. The vertical dark line at 1:5 points out the case shown in Fig.~\ref{Fig2} a), i.~e., the 1:5 aspect-ratio oblate case. This panel shows that the MBP expression approaches better the exact case for all studied aspect-ratios --from the 1:25 oblate (leftmost points) to the 25:1 prolate (rightmost points)-- and for both, $\gamma=1$ and $\gamma=\gamma_{min}$. The only exception is the 1:1 ellipsoid (sphere) where all approaches yield the exact result. Prolate ellipsoids, however, are reasonably described by BP for all aspect-ratios, and thus, the gain by implementing MBP in these cases is not very large. Conversely, the main BP deviations occur for oblate shapes reaching $E$ values over $10\%$ for the 1:20 and 1:25 cases. In these cases the gain of MBP is remarkable, since it always shows $E$ values below $4\%$. In view of these results, it seems that the gain in precision is worth the little extra operations needed to compute MBP instead of BP. In addition, the gain of employing $\gamma=\gamma_{min}$ instead of $\gamma=1$ is close to $50\%$ (see Fig.~\ref{Fig2} b)). The values of $\gamma_{min}$ are shown in Fig.~\ref{Fig2} b) as open symbols. For 1:5 oblates $\gamma_{min}=0.794$.

\section{Replica Exchange Monte Carlo}

Even though the analytical expressions for determining whether or not two ellipsoids overlap are relatively fast to compute, sampling from crowded systems is always a difficult task~\cite{Odriozola11}. We expect 1:5 oblate hard ellipsoids to show a fluid-nematic transition at relatively low densities~\cite{Frenkel84,Frenkel85,Samborski94} and a nematic-crystal transition at high densities~\cite{Frenkel84,Frenkel85}, and this work attempts to capture both. By analogy with the hard sphere case, we also expect the crystal equilibrium branch of the phase diagram to lie below a metastable nematic phase branch (the continuation of the nematic branch toward larger densities than the crystallization point), which surely hinders equilibrating the system at high densities~\cite{Odriozola09}. For these reasons, we are implementing the replica exchange Monte Carlo methodology, which is well proven to assist the systems to reach equilibrium at difficult (high density / low temperature) conditions.     

In the classical replica exchange scheme, $n_r$ identical replicas are considered each following a typical canonical
simulation at different temperatures~\cite{Marinari92,Lyubartsev92,hukushima96}. Thus, an extended ensemble can be defined so that its partition function is $Q_{extended}=\prod_{i=1}^{n_r}Q_{NVTi}$, being $Q_{NVTi}$ the partition function of ensemble $i$ at
temperature $T_i$, number of ellipsoids $N$, and volume $V$. The existence of this extended ensemble justifies the introduction of
swap trial moves between any two ensembles (each being sampled by only one replica at a time), whenever the detail
balance condition is satisfied. If all $(i,T_i)(j,T_j)\!\!\rightarrow\!(j,T_i)(i,T_j)$ swap trials have
the same a priori probability of being performed, the swap acceptance probability becomes
\begin{equation}\label{accT}
P_{acc}\!=\! min(1,\exp[( \beta_j - \beta_i)(U_{i}-U_{j})])
\end{equation}
where $\beta_i=1/(k_BT_i)$ is the reciprocal temperature of replica $i$, $k_B$ is the Boltzmann's constant, and $U_i$ is the
energy of replica $i$. Hence, by introducing these swap trials, a particular replica travels through many temperatures allowing it to overcome free-energy barriers. Additionally, sampling on particular ensembles is not disturbed but enriched by the different contributions of the $n_r$ replicas. 

Since we are dealing with hard ellipsoids the temperature plays a trivial role. Then, to take advantage of the method, we must perform the expansion in pressure instead of temperature~\cite{Odriozola09}. Hence, the partition function of the extended ensemble is given by \cite{Okabe01,Odriozola09} 
\begin{equation}
Q_{\rm extended}=\prod_{i=1}^{n_r} Q_{N T P_i},
\end{equation} 
where $Q_{NTP_i}$ is the partition function of the isobaric-isothermal ensemble of the system at pressure $P_i$, temperature $T$, and with $N$ particles. 

This extended ensemble is sampled by combining standard $NTP_i$ simulations on each replica (involving trial displacements,  rotations of single ellipsoids, and trial volume changes) and replica exchanges (swap moves at the replica level). To satisfy detailed balance, these swap moves are performed by setting equal all a priory probabilities for choosing adjacent pairs of replicas and using the following acceptance probability~\cite{Odriozola09}
\begin{equation}
\label{accP} 
P_{\rm acc}\!=\! \min(1,\exp[\beta(P_i-P_j)(V_i-V_j)]), 
\end{equation} 
where $V_i-V_j$ is the volume difference between replicas $i$ and $j$. Adjacent pressures should be close enough to provide reasonable exchange acceptance rates between neighboring ensembles. In order to take good advantage of the method, the ensemble at the smaller pressure must also ensure large jumps in configuration space, so that the larger pressure ensembles can be efficiently sampled.

The probability for selecting an ellipsoid displacement trial, $P_d$, an ellipsoid rotation, $P_r$, for selecting a volume change trial, $P_v$, and a swap trial, $P_s$, are fixed to  
\begin{equation}
\label{PdPvPs} 
\begin{array}{lll} 
P_d & = & P_r = n_rN/(n_r(2N+1)+w), \\
P_v & = & n_r/(n_r(2N+1)+w), \\
P_s & = & w/(n_r(2N+1)+w), \\ 
\end{array} 
\end{equation} 
where $w \ll 1$ is a weight factor. Note that $P_d+P_r+P_v+P_s=1$, as it should. The probability density function to have the next swap trial move at the trial $n_t$ is given by
\begin{equation}
\label{Pdfs} P(n_t) = P_s \exp(-P_s n_t). 
\end{equation}
Hence, we may obtain the next swap trial move from $n_t= - \ln(\xi) / P_s$, with $\xi$ being a random number uniformly distributed
in the interval $(0,1)$ \cite{Gillespie77,Odriozola03}. We set all ellipsoids of a given replica to have the same a priori probability of being selected to perform a displacement or a rotational trial. The same is true for selecting a replica to perform a volume change trial.

The trials $[1,n_t-1]$ are displacements, rotations, and volume changes, and so, they can be independently performed on the replicas. This has the advantage of being easily parallelized. The algorithm is parallelized through message passing interface (mpi) fortran in $n_r$ threads, though quad core desktops are used. Since all swap trials are performed in a single thread, the efficiency
of the parallelization increases with decreasing $w$. We employed $w=1/100$. Verlet neighbor lists~\cite{Donev05a,Donev05b} are used for saving CPU time, which can be quite large for the replicas evolving with the highest pressure values. These lists must take into account displacements and rotations to timely update.  
 
Simulations are performed in two steps. All simulations are started by randomly placing the ellipsoids in a random orientation (avoiding overlaps), so that the initial volume fraction is $\varphi= v_{e} \rho = 0.2$, where $\rho$ is the number density, 
$v_{e}=4 \pi \sigma_{\|} \sigma_{\bot}^2/3$ is the ellipsoid volume (for all models), $\sigma_{\|}=1$, and $\sigma_{\bot}=5$. We first perform about $2 \times 10^{13}$ trial moves at the desired state points, during which we observe that the replicas reach a stationary state (equilibrating procedure). We then perform $2 \times 10^{13}$ additional trials during which various measurements are carried out, with results described in the following section.

The maximum particle displacements, maximum rotational displacements, and volume changes for trial moves are adapted for each pressure to yield acceptance rates close to 0.3. Thus, particle displacements, rotations, and volume changes of ensembles having high pressures are smaller than those associated to ensembles having low pressures. An optimal allocation of the replicas should lead to a constant swap acceptance rate for all pairs of adjacent ensembles. For a temperature expansion, the efficiency of the method peaks at swap acceptance rates close to 20\% \cite{Rathore05}. In this regard we implemented a simple algorithm to smoothly adjust the state points (pressures) to yield an approximately constant swap acceptance rate while keeping constant the maximum and minimum pressure. At the starting point, we use a geometric progression of the pressure with the replica index. Note that the adaptation of maximum displacements and the shift of the desired pressures violates the detail balance condition. Thus, these procedures are performed only during the equilibrating procedure (the first $2 \times 10^{13}$ trial moves). This work is performed by considering $N=100$ and $N=200$ ellipsoids, and $n_r=64$ (to cover a wide range of densities while keeping swap acceptance rates over 20\%).

\section{Results}
Results are split in three subsections. The first one is devoted to compare the equation of state obtained by considering the exact model of the 1:5 oblate ellipsoids with the analytical expression of Berne and Pechukas (BP) and its modification (MBP) presented in subsection~\ref{subsecMBP}. This is done by considering 100 particles (N=100). The second subsection compares the N=100 MBP case with $\gamma=\gamma_{min}$ to simulations performed with N=200. The obtained structures are analyzed in a third subsection for this larger system size. 

\subsection{Exact vs Analytical overlap distance}

\begin{figure}
\resizebox{0.46\textwidth}{!}{\includegraphics{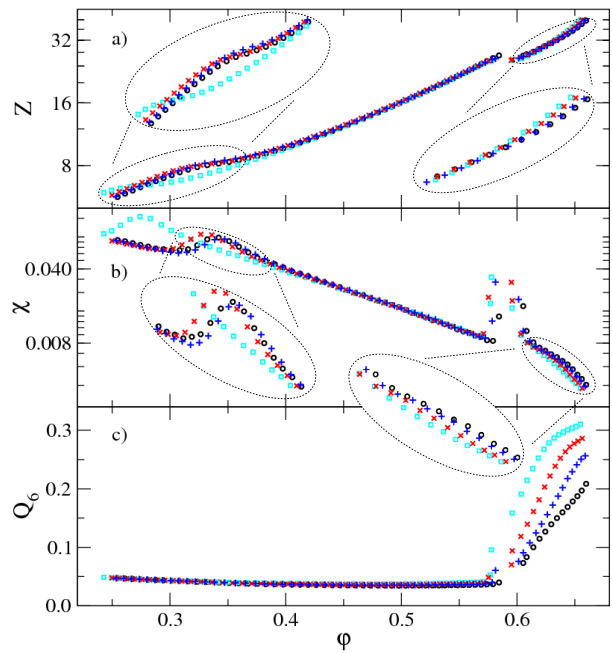}}
\caption{\label{Fig3} a) Equations of state of the exact and analytical hard 1:5 oblate ellipsoidal models, $Z(\varphi)$ ($\varphi$ is the volume fraction). b) Isothermal compressibility obtained from density fluctuations, $\chi(\varphi)$. c) Order parameter, $Q_6(\varphi)$. For all panels, black circles, cyan squares, red crosses, and blue plus symbols correspond to the exact PW overlap distance, and the analytical solutions given by BP, MBP with $\gamma=1$, and MBP with $\gamma=\gamma_{min}$, respectively. All data correspond to $N=100$.}
\end{figure}

The exact hard 1:5 oblate ellipsoidal model can be studied for a moderate system size. This is possible since, on the one hand, the exact iterative procedure to solve the $\sigma_{PW}$ overlap distance is relatively fast, and on the other, the numerical solution is computed only for those few cases where the ellipsoids geometric centers distance, $r$, is less than $\sigma_{BP}$. In this way the exact PW simulation is approximately three times more computationally demanding than the analytical cases. As mentioned, $n_r=64$ replicas are employed to cover a wide pressure range while keeping large swap acceptance rates. The probability distribution functions (PDFs) obtained for this system are given in previous work~\cite{OdriozolaHE}. From them the dimensionless pressure $Z=\beta P/\rho$ and the isothermal compressibility $\chi$ are calculated as a function of the most frequent volume fraction, $\varphi$. These functions are given in panels a) and b) of Fig.~\ref{Fig3} as black circles. The $\chi$ values are obtained by means of the density fluctuations, i.~e., by $\chi=N(<\rho^2>-<\rho>^2)/<\rho>^2$, which should equal $\chi= \delta \rho / \delta (\beta P)$ according to the fluctuation-dissipation theorem. Finally, panel c) shows the order parameter $Q_6=\left(\frac{4\pi}{13}\sum_{m=-6}^{m=6}|<\!Y_{6m}(\theta,\phi)\!>|^2\right)^{1/2}$, being $<\!Y_{6m}(\theta, \phi)\!>$ the average over all bonds and configurations of the spherical harmonics of the orientation polar angles $\theta$ and $\phi$~\cite{Steinhardt96,Rintoul96b,Odriozola09}. $Q_6$ approaches zero for a completely random system of a large number of points,  and increases when configurations present angular order. All these data shown as black circles correspond to the exact solution and are taken from ref.~\cite{OdriozolaHE}. Also from this reference, we are including the data corresponding to the BP overlap distance approximation as cyan squares.

Simulations for the exact PW case show a isotropic-nematic transition at $\varphi=0.341$ and a nematic-crystal transition at the volume fraction interval of $0.584-0.605$~\cite{OdriozolaHE}. The fluid-fluid transition is evidenced by a plateau of $Z(\varphi)$, a kink of $\chi(\varphi)$, and a practically invariant $Q_6(\varphi)$. The fluid-crystal transition is characterized by a jump of $\varphi$ accompanied with a kink of $\chi(\varphi)$ and a steep increase of $Q_6(\varphi)$. For the isotropic region $Z(\varphi)$ agrees with the simulation data provided by Mc.~Bride and Lomba~\cite{McBride07} and so, our $Z(\varphi)$ is described by the Vega equation of state for the isotropic fluid~\cite{Vega97,McBride07}. The BP model also captures the three phases, although the transitions are shifted to lower densities, and the low and high density branches differ from the exact case (see the insets of Fig.~\ref{Fig3}). The transitions occur at $\varphi=0.274$ and at the range $0.577-0.595$. These discrepancies are due to the fact that $\sigma_{BP}\geq\sigma_{PW}$. For the nematic phase, however, a very good agreement between BP and PW is seen for all panels of Fig.~\ref{Fig3}. This is due to $\sigma_{BP}$ well approaches $\sigma_{PW}$ for parallel configurations (see Fig.~\ref{Fig1}). Discrepancies are more pronounced at low densities where T-shape configurations are frequent, for which $\sigma_{BP}$ may reach $1.2 \times \sigma_{PW}$ (see Fig.~\ref{Fig2}). Thus, the MBP approach, which corrects the T-shape BP mismatch, is expected to approach better the PW simulations.       

The MBP data with $\gamma=1$ are shown as red crosses in the three panels and the insets of Fig.~\ref{Fig3}. As can be seen, discrepancies between the exact and the analytical cases are clearly diminished. In particular, the isotropic-nematic transition is now observed at $\varphi=0.331$, and so, the relative difference with the exact case diminishes from 20$\%$ to 3$\%$. Notwithstanding, the exact-analytical agreement can be further improved by setting $\gamma=\gamma_{min}$. The corresponding results are also included in Fig.~\ref{Fig3} as blue plus symbols. In this case all discrepancies practically vanish (though a somewhat more structured crystal is still produced according to $Q_6$). The obtained transitions occur at $\varphi=0.344$, and in the range $0.581-0.599$.

\subsection{Size effects}

\begin{figure}
\resizebox{0.45\textwidth}{!}{\includegraphics{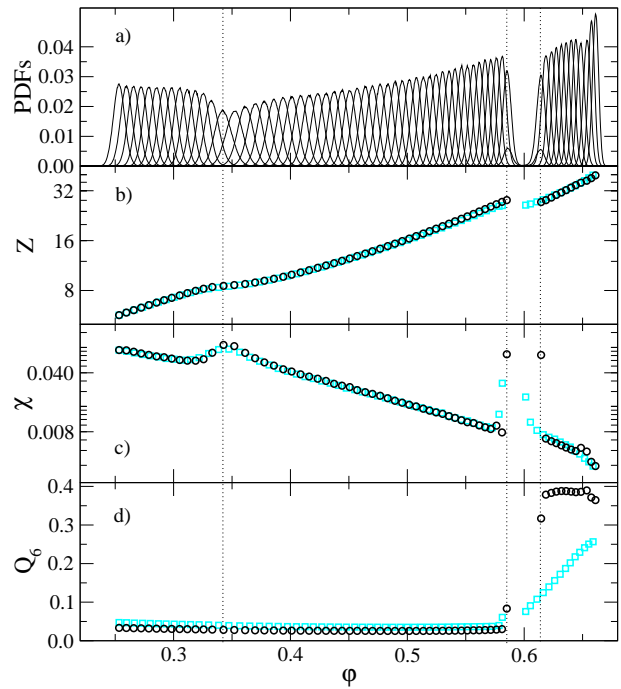}}
\caption{\label{Fig4} a) Probability distribution functions (PDFs) of volume fraction fluctuations for each of the $n_r=64$ pressure values and for $N=200$. b) Equations of state, $Z(\varphi)$. c) Isothermal compressibilities, $\chi(\varphi)$, obtained from density fluctuations. d) Order parameters, $Q_6(\varphi)$. For panels b), c), and d), dark circles and light squares correspond to $N=200$ and $N=100$, respectively. Vertical dotted lines highlight the $N=200$ transitions. All data correspond to the analytical MBP case with $\gamma=\gamma_{min}$.}
\end{figure}

The faster computation of the analytical case allows us to study a larger system size, i.~e., $N=200$. For this system size and keeping $n_r=64$ we obtain swap acceptance rates close to $30\%$. These high acceptance rates are a consequence of the large overlapping areas of the probability distribution functions (PDFs) shown in panel a) of Fig.~\ref{Fig4}. There, the overall trend of the PDFs to get narrower and higher with increasing density (pressure) is seen. This makes a general decrease of the isothermal compressibility $\chi= \delta \rho / \delta (\beta P)$ evident. The overall trend is disrupted at the transitions where PDFs are distorted from their natural Gaussian shape. Indeed, PDFs turn bimodal at the nematic-crystal transition. In brief, all features captured by the $N=100$ system also appear for the $N=200$ case (panel a) should be compared to Fig.~1 of previous work~\cite{OdriozolaHE}). Nonetheless, some differences appear: The PDFs are higher and narrower due to the larger system size, the nematic-crystal transition density gap enlarges and shifts to larger densities, and a small disruption of the PDFs heights trend appears at very large $\varphi$ ($\varphi \approxeq 0.65$). In particular, the nematic-crystal transition shift is clearly shown in panels b), c), and e) of the same figure, where the $N=100$ data are included as cyan squares for an easy comparison. This is the only practical difference of $Z(\varphi)$. $\chi(\varphi)$ evidences that size effects are absent from the isotropic-nematic transition (this is in agreement with Allen and Mason results for the 3:1 prolate case~\cite{Allen95} and opposed to the Zarragoicoechea et al.~suggestion~\cite{Zarragoicoechea92}). On the other hand, it reflects the small disruption of the PDFs heights trend as a little kink at high densities. Finally, $Q_6(\varphi)$ suggests the formation of less defective crystals for the larger system size ($Q_6$ is larger for $N=200$ at densities above the nematic-crystal transition).   

For $N=200$, the isotropic-nematic transition occurs at $\varphi=0.345$ and $Z=8.6$. The nematic-solid transition is located in the volume fraction interval $0.585-0.614$ for $Z=27.7$. Thus, we can now extrapolate the values corresponding to $N=100$ and 200 to $N\rightarrow \infty$. This procedure leads to $\varphi=0.346$ and $Z=8.7$ for the isotropic-nematic transition and to $\varphi$ in the range of $0.589-0.629$ at $Z=29.4$ for the nematic-solid transition, which are our estimates for the MBP model at the thermodynamic limit. Since a similar shift is expected for the exact case, our estimate for the thermodynamic limit for the isotropic-nematic transition is $\varphi= 0.343 \pm 0.003$ and $Z=8.4 \pm 0.3$, and in the volume fraction interval $(0.592 \pm 0.006)-(0.634 \pm 0.008)$ at $Z=30.4 \pm 0.9$ for the nematic-solid transition. These values can be compared to those shown by the phase diagrams given in refs.~\cite{Frenkel84,Frenkel85}. We estimated them to be $\varphi \approxeq 0.37$ for the isotropic-nematic transition and the interval $\varphi \approxeq 0.60-0.67$ for the nematic-solid transition. Additionally, Samborski et.~al. placed the isotropic-nematic transition at the $\varphi$ range of $0.333-0.351$~\cite{Samborski94}. Hence, our value (taken where function $\chi(\varphi)$ peaks) lies exactly in the middle of this range. We should also add that replicas with $\varphi \approxeq 0.343$ are not totally nematic or isotropic according to the observed snapshots (not shown). Thus, the transition takes place over a volume fraction range instead of a single point. Taking the two intersection points of $\chi(\varphi)$ with the horizontal line that contains the local minimum of $\chi(\varphi)$ (at $\varphi=0.312$ for $N=100$ and the exact model) we obtain the volume fraction range $0.312-0.363$. This range translates into $0.314-0.365$ by applying the corresponding shift to estimate the transition at the thermodynamic limit.

\subsection{Structure}

\begin{figure}
\resizebox{0.45\textwidth}{!}{\includegraphics{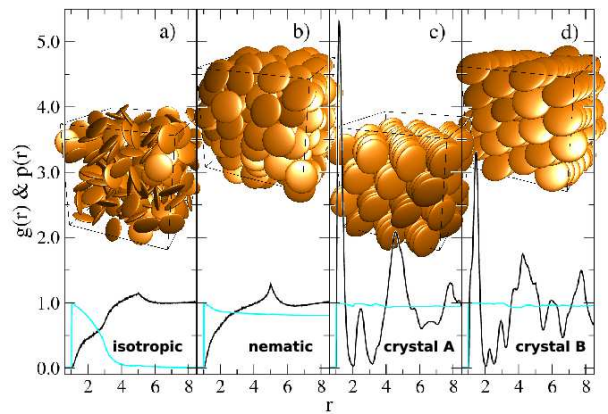}}
\caption{\label{Fig5} Radial distribution functions (black lines), $g(r)$, and their corresponding order parameter (cyan lines), $p(r)$, for different pressures (increasing from left to right). Panels a), b), c), and d) correspond to the isotropic, nematic, crystal A, and crystal B structures. The insets show the corresponding snapshots. }
\end{figure}

Fig.~\ref{Fig5} shows the radial distribution functions, $g(r)$, and order parameter, $p(r)=<1/2(3(\hat{\mathbf{u}}_{i} \cdot \hat{\mathbf{u}}_{j})^2-1)>$, for different pressures. Panel a) corresponds to the lowest pressure where both radial functions signal an isotropic phase. That is, the $g(r)$ smoothly increases peaking at $r\approxeq \sigma_{\bot}$, while the $p(r)$ shows no sign of long range angular order at large distances (the $p(r)$ peak at $r = \sigma_{\|}$ results from the few and forced parallel configurations at short distances). The corresponding snapshot clearly shows this fact. Panel b) corresponds to the structure found for $\varphi\approxeq 0.45$. In this case the function $p(r)$ shows a long range alignment of the ellipsoids though the $g(r)$ still points to a fluid phase (nematic). The only peak shown by the $g(r)$, as in the previous case, corresponds to $r\approxeq \sigma_{\bot}$, but now the peak is higher probably due to the side-to-side configurations in the developed layers (see the corresponding inset). The last two panels correspond to a solid phase. Pressure in c) is close to the transition and pressure in d) is the highest. Both crystals differ from each other. The structure of panel c) (crystal A) is the one found for the $N=100$ case of previous work~\cite{OdriozolaHE}. It is characterized by a first and large peak at $r\approxeq 1.2$, which corresponds to the touching and stacked ellipsoids shown in the inset. There is a second peak at $r\approxeq 2.4$, corresponding to two ellipsoids separated by a third one and sandwiched by them (all belonging to the same stack). Finally, the wide peak at $r\approxeq 4.3$ corresponds to side-to-side configurations of particles belonging to different stacks. This peak is smaller than $\sigma_{\bot}$ since the particles of the adjacent stacks are partially sandwiched. The structure of panel d) (crystal B) shows a shorter main peak which is also shifted to $r\approxeq 1.5$ (the shoulder at $r\approxeq 1.2$ and the small peak at $r\approxeq 2.4$ correspond to contributions to the average from replicas having a crystal A like structure). In this arrangement, the secondary peak appearing at $r\approxeq 2.4$ is also shifted to $r\approxeq 3.0$. Thus, in both cases, the second peak appears at a distance two times larger than that of the main peak. Other differences appear at larger distances which are related to the way the stacks are arranged. 

\begin{figure}
\resizebox{0.45\textwidth}{!}{\includegraphics{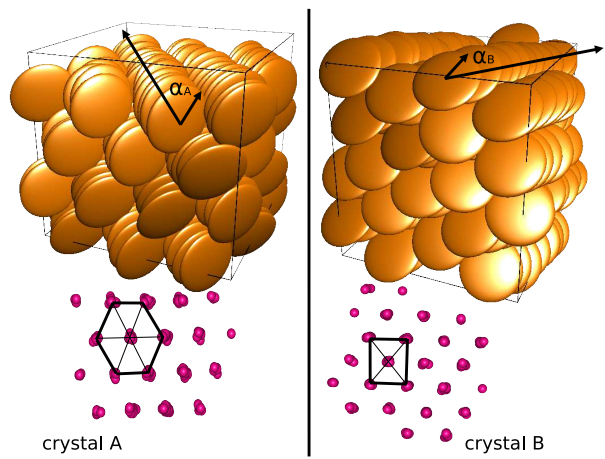}}
\caption{\label{Fig6} Detail of the different crystal structures found in the $N=200$ system (cases d) and e) of Fig.~\ref{Fig4}). The shown $\alpha$ angles, defined as $<90^o-\cos^{-1}(\hat{\mathbf{u}}_i \cdot \hat{\mathbf{r}}_s)>$ where $\hat{\mathbf{r}}_s$ is the versor pointing along the stack axis, are close to $62^o$ and $44^o$ for crystal A and B, respectively. }
\end{figure}

The differences found for the radial distribution functions can be understood from the structural details shown in figure~\ref{Fig6}. From this figure it is seen that the projection of the particles geometric centers towards a plane perpendicular to the stacks axes are arranged in a hexagonal lattice for crystal A (left panel), whereas they produce a square centered pattern for crystal B (right panel). Additionally, the intra-stack arrangements also differ respect to the other. In crystal A the entities of a given stack are more perpendicular to the stack axis than for the crystal B structure. In other words, the angles $\alpha=<90^o-\cos^{-1}(\hat{\mathbf{u}}_i \cdot \hat{\mathbf{r}}_s)>$ are $62^o$ and $44^o$ for crystal A and crystal B, respectively. The more tilted intra-stack arrangement of crystal B leads to an increase of the number of stacks to be arranged for a given area (perpendicular to the stacks' axes), though the stacks contain less particles for a given axis depth. This more tilted intra-stack arrangement explains the shift of the first two peaks of the $g(r)$ towards larger distances of crystal B.

Note that crystal B is not detected for a system size of $N=100$. Additionally, the crystal A structure found for $N=100$ shows many more defects than the one found for $N=200$. This is evidenced by the smaller $Q_6$ values and also directly seen from the snapshots. Hence, size effects are important for determining the correct crystal phase (implementing non-orthogonal unit cells by allowing the lattice vectors to change may improve this point). Accordingly, we can only state that the structures here reported correspond to the studied system size. For $N=200$, we observe that replicas having a structure like crystal B are preferably located at higher pressures, though more replicas produce the crystal A structure. Consequently, a kink appears in $\chi(\varphi)$ at high pressures, separating a rich crystal A region at smaller pressures from a rich crystal B region at higher pressures. This points out that crystal B has a larger compressibility than crystal A and may suggest a crystal-crystal transition. Finally, we should also mention that the structures here reported are different from those pointed out by Donev et.~al.~\cite{Donev04a,Donev04b}. They reported that structures composed by a two-layer lamination where layers are disposed rotated $\pi/2$ with respect to the other yields a packing fraction of $\varphi=0.7707$. To obtain this large packing fraction layers must be face-centered square planar and have all ellipsoids oriented with one of their semiaxes perpendicular to the layer and the other two oriented along the axes of the lattice. The authors also mention that there is nothing suggesting that this family of structures is the densest. On the other hand, a fit of the form $Z \sim (\varphi_d-\varphi)^{-1}$ to the high pressure branch of $Z(\varphi)$ for $N=200$ (panel b) of Fig.~\ref{Fig4}) leads to a divergence at $\varphi_d=0.769\pm 0.002$, which is virtually equal to $\varphi=0.7707$. Consequently, several crystal candidates for the equilibrium structure at high densities seem to exist.

\section{Conclusions}

We used the replica exchange Monte Carlo technique to produce the equation of state of hard 1:5 aspect-ratio oblate ellipsoids for a wide density range. In addition to the analytical approximation of the overlap distance given by Berne and Pechukas (BP) and the exact numerical solution given by Perram and Wertheim, we implemented a simple modification of the original BP approximation, which corrects the known T-shape mismatch of BP. We found that this approximation produces an equation of state practically equal to that obtained for the exact overlap distance solution. We then used this approach to study a larger system size ($N=200$). The produced results allowed us to estimate the locations of the isotropic-nematic and nematic-crystal transitions for the thermodynamic limit at $\varphi= 0.343 \pm 0.003$ and $Z=8.4 \pm 0.3$ and in the interval $(0.592 \pm 0.006)-(0.634 \pm 0.008)$ for $Z=30.4 \pm 0.9$, respectively.  

\begin{acknowledgments}
The authors thank projects Nos.~Y.00116 and Y.00119 SENER-CONACyT for financial support. Authors also thank Tito Gonz\'alez Rodr\'{\i}guez for reading the manuscript.   
\end{acknowledgments}


%

\end{document}